


%
%


\def\famname{
 \textfont0=\textrm \scriptfont0=\scriptrm
 \scriptscriptfont0=\sscriptrm
 \textfont1=\textmi \scriptfont1=\scriptmi
 \scriptscriptfont1=\sscriptmi
 \textfont2=\textsy \scriptfont2=\scriptsy \scriptscriptfont2=\sscriptsy
 \textfont3=\textex \scriptfont3=\textex \scriptscriptfont3=\textex
 \textfont4=\textbf \scriptfont4=\scriptbf \scriptscriptfont4=\sscriptbf
 \skewchar\textmi='177 \skewchar\scriptmi='177
 \skewchar\sscriptmi='177
 \skewchar\textsy='60 \skewchar\scriptsy='60
 \skewchar\sscriptsy='60
 \def\rm{\fam0 \textrm} \def\bf{\fam4 \textbf}}
\def\sca#1{scaled\magstep#1} \def\scah{scaled\magstephalf} 
\def\twelvepoint{
 \font\textrm=cmr12 \font\scriptrm=cmr8 \font\sscriptrm=cmr6
 \font\textmi=cmmi12 \font\scriptmi=cmmi8 \font\sscriptmi=cmmi6 
 \font\textsy=cmsy10 \sca1 \font\scriptsy=cmsy8
 \font\sscriptsy=cmsy6
 \font\textex=cmex10 \sca1
 \font\textbf=cmbx12 \font\scriptbf=cmbx8 \font\sscriptbf=cmbx6
 \font\it=cmti12
 \font\sectfont=cmbx12 \sca1
 \font\refrm=cmr10 \scah \font\refit=cmti10 \scah
 \font\refbf=cmbx10 \scah
 \def\twelverm{\textrm} \def\twelveit{\it} \def\twelvebf{\textbf}
 \famname \textrm 
 \voffset=.04in \hoffset=.21in
 \normalbaselineskip=18pt plus 1pt \baselineskip=\normalbaselineskip
 \parindent=21pt
 \setbox\strutbox=\hbox{\vrule height10.5pt depth4pt width0pt}}


\catcode`@=11

{\catcode`\'=\active \def'{{}^\bgroup\prim@s}}

\def\screwcount{\alloc@0\count\countdef\insc@unt}   
\def\screwdimen{\alloc@1\dimen\dimendef\insc@unt} 
\def\screwbox{\alloc@4\box\chardef\insc@unt}

\catcode`@=12


\overfullrule=0pt			
\voffset=.04in \hoffset=.21in
\vsize=9in \hsize=6in
\parskip=\medskipamount	
\lineskip=0pt				
\normalbaselineskip=18pt plus 1pt \baselineskip=\normalbaselineskip
\abovedisplayskip=1.2em plus.3em minus.9em 
\belowdisplayskip=1.2em plus.3em minus.9em	
\abovedisplayshortskip=0em plus.3em	
\belowdisplayshortskip=.7em plus.3em minus.4em	
\parindent=21pt
\setbox\strutbox=\hbox{\vrule height10.5pt depth4pt width0pt}
\def\makefootline{\baselineskip=30pt \line{\the\footline}}
\footline={\ifnum\count0=1 \hfil \else\hss\twelverm\folio\hss \fi}
\pageno=1


\def\boxit#1{\leavevmode\thinspace\hbox{\vrule\vtop{\vbox{
	\hrule\kern1pt\hbox{\vphantom{\bf/}\thinspace{\bf#1}\thinspace}}
	\kern1pt\hrule}\vrule}\thinspace}
\def\Boxit#1{\noindent\vbox{\hrule\hbox{\vrule\kern3pt\vbox{
	\advance\hsize-7pt\vskip-\parskip\kern3pt\bf#1
	\hbox{\vrule height0pt depth\dp\strutbox width0pt}
	\kern3pt}\kern3pt\vrule}\hrule}}


\def\put(#1,#2)#3{\screwdimen\unit  \unit=1in
	\vbox to0pt{\kern-#2\unit\hbox{\kern#1\unit
	\vbox{#3}}\vss}\nointerlineskip}

%
%
%
%
%
%
%

\def\\{\hfil\break}

\def\center{\leftskip=0pt plus 1fill \rightskip=\leftskip \parindent=0pt
 \def\textindent##1{\par\hangindent21pt\footrm\noindent\hskip21pt
 \llap{##1\enspace}\ignorespaces}\par}
\def\unnarrower{\leftskip=0pt \rightskip=\leftskip}
\def\thetitle#1#2#3#4#5{
 \font\titlefont=cmbx12 \sca2 \font\footrm=cmr10 \font\footit=cmti10
  \twelverm
	{\hbox to\hsize{#4 \hfill ITP-SB-#3}}\par
	\vskip.8in minus.1in {\center\baselineskip=1.44\normalbaselineskip
 {\titlefont #1}\par}{\center\baselineskip=\normalbaselineskip
 \vskip.5in minus.2in #2
	\vskip1.4in minus1.2in {\twelvebf ABSTRACT}\par}
 \vskip.1in\par
 \narrower\par#5\par\unnarrower\vskip3.5in minus2.3in\eject}
\def\paper\par#1\par#2\par#3\par#4\par#5\par{\twelvepoint
	\thetitle{#1}{#2}{#3}{#4}{#5}} 
\def\author#1#2{#1 \vskip.1in {\twelveit #2}\vskip.1in}
\def\ITP{Institute for Theoretical Physics\\
	State University of New York, Stony Brook, NY 11794-3840}
\def\WS{W. Siegel\footnote{${}^1$}{       
 Internet address: siegel@insti.physics.sunysb.edu.}}


\def\sect#1\par{\par\ifdim\lastskip<\medskipamount
	\bigskip\medskip\goodbreak\else\nobreak\fi
	\noindent{\sectfont{#1}}\par\nobreak\medskip} 
\def\itemize#1 {\item{[#1]}}	
\def\vol#1 {{\refbf#1} }		 

\def\ref#1{\setbox0=\hbox{M}$\vbox to\ht0{}^{#1}$}


\def\NP #1 {{\refit Nucl. Phys.} {\refbf B{#1}} }
\def\PL #1 {{\refit Phys. Lett.} {\refbf{#1}} }
\def\PR #1 {{\refit Phys. Rev. Lett.} {\refbf{#1}} }
\def\PRD #1 {{\refit Phys. Rev.} {\refbf D{#1}} }


\hyphenation{pre-print}
\hyphenation{quan-ti-za-tion}


\def\watch{
 \screwcount\hrs
 \screwcount\mins
 \screwcount\merid
 \screwcount\hrmins
 \screwcount\hrmerid
 \hrs=\time
 \mins=\time
 \divide\hrs by 60
 \merid=\hrs
 \hrmins=\hrs
 \divide\merid by 12
 \hrmerid=\merid
 \multiply\hrmerid by 12
 \advance\hrs by -\hrmerid
 \ifnum\hrs=0\hrs=12\fi
 \multiply\hrmins by 60
 \advance\mins by -\hrmins
 \number\hrs:\ifnum\mins<10 {0}\fi\number\mins\space
 \ifnum\merid=0 AM\else PM\fi}
\def\today{\ifcase\month\or
 January\or February\or March\or April\or May\or June\or July\or
 August\or September\or October\or November\or December\fi
 \space\number\day, \number\year}

%
%

\def\on#1#2{{\buildrel{\mkern2.5mu#1\mkern-2.5mu}\over{#2}}}
\def\dt#1{\on{\hbox{\bf .}}{#1}}                
\def\ddt#1{\on{\hbox{\bf .\kern-1pt.}}#1}    
\def\slap#1#2{\setbox0=\hbox{$#1{#2}$}
	#2\kern-\wd0{\hbox to\wd0{\hfil$#1{/}$\hfil}}}
\def\sla#1{\mathpalette\slap{#1}}                
\def\bop#1{\setbox0=\hbox{$#1M$}\mkern1.5mu
	\vbox{\hrule height0pt depth.04\ht0
	\hbox{\vrule width.04\ht0 height.9\ht0 \kern.9\ht0
	\vrule width.04\ht0}\hrule height.04\ht0}\mkern1.5mu}
\def\bo{{\mathpalette\bop{}}}                        
\def~{\widetilde} 
\mathcode`\*="702A                  
\def\in{\relax\ifmmode\mathchar"3232\else{\refit in\/}\fi} 
\def\f#1#2{{\textstyle{#1\over#2}}}	   
\def\half{{\textstyle{1\over{\raise.1ex\hbox{$\scriptstyle{2}$}}}}}

\catcode`\^^?=13				    
\catcode128=13 \def €{\"A}                 
\catcode129=13 \def {\AA}                 
\catcode130=13 \def '{\c}           	   
\catcode131=13 \def ƒ{\'E}                   
\catcode132=13 \def "{\~N}                   
\catcode133=13 \def …{\"O}                 
\catcode134=13 \def †{\"U}                  
\catcode135=13 \def ‡{\'a}                  
\catcode136=13 \def ˆ{\`a}                   
\catcode137=13 \def ‰{\^a}                 
\catcode138=13 \def Š{\"a}                 
\catcode139=13 \def ‹{\~a}                   
\catcode140=13 \def Œ{\alpha}            
\catcode141=13 \def {\chi}                
\catcode142=13 \def Ž{\'e}                   
\catcode143=13 \def {\`e}                    
\catcode144=13 \def {\^e}                  
\catcode145=13 \def '{\"e}                
\catcode146=13 \def '{\'\i}                 
\catcode147=13 \def "{\`\i}                  
\catcode148=13 \def "{\^\i}                
\catcode149=13 \def •{\"\i}                
\catcode150=13 \def –{\~n}                  
\catcode151=13 \def —{\'o}                 
\catcode152=13 \def ˜{\`o}                  
\catcode153=13 \def ™{\^o}                
\catcode154=13 \def š{\"o}                 
\catcode155=13 \def ›{\~o}                  
\catcode156=13 \def œ{\'u}                  
\catcode157=13 \def {\`u}                  
\catcode158=13 \def ž{\^u}                
\catcode159=13 \def Ÿ{\"u}                
\catcode160=13 \def  {\tau}               
\catcode161=13 \mathchardef ¡="2203     
\catcode162=13 \def ¢{\oplus}           
\catcode163=13 \def £{\relax\ifmmode\to\else\itemize\fi} 
\catcode164=13 \def ¤{\subset}	  
\catcode165=13 \def ¥{\infty}           
\catcode166=13 \def ¦{\mp}                
\catcode167=13 \def §{\sigma}           
\catcode168=13 \def ¨{\rho}               
\catcode169=13 \def ©{\gamma}         
\catcode170=13 \def ª{\leftrightarrow} 
\catcode171=13 \def «{\relax\ifmmode\acute\else\expandafter\'\fi}
\catcode172=13 \def ¬{\relax\ifmmode\expandafter\ddt\else\expandafter\"\fi}
\catcode173=13 \def ­{\equiv}            
\catcode174=13 \def ®{\approx}          
\catcode175=13 \def ¯{\Omega}          
\catcode176=13 \def °{\otimes}          
\catcode177=13 \def ±{\ne}                 
\catcode178=13 \def ²{\le}                   
\catcode179=13 \def ³{\ge}                  
\catcode180=13 \def ´{\upsilon}          
\catcode181=13 \def µ{\mu}                
\catcode182=13 \def ¶{\delta}             
\catcode183=13 \def ·{\epsilon}          
\catcode184=13 \def ¸{\Pi}                  
\catcode185=13 \def ¹{\pi}                  
\catcode186=13 \def º{\beta}               
\catcode187=13 \def »{\partial}           
\catcode188=13 \def ¼{\nobreak\ }       
\catcode189=13 \def ½{\zeta}               
\catcode190=13 \def ¾{\sim}                 
\catcode191=13 \def ¿{\omega}           
\catcode192=13 \def À{\dt}                     
\catcode193=13 \def Á{\gets}                
\catcode194=13 \def Â{\lambda}           
\catcode195=13 \def Ã{\nu}                   
\catcode196=13 \def Ä{\phi}                  
\catcode197=13 \def Å{\xi}                     
\catcode198=13 \def Æ{\psi}                  
\catcode199=13 \def Ç{\int}                    
\catcode200=13 \def È{\oint}                 
\catcode201=13 \def É{\relax\ifmmode\cdot\else\vol\fi}    
\catcode202=13 \def Ê{\relax\ifmmode\,\else\thinspace\fi}
\catcode203=13 \def Ë{\`A}                      
\catcode204=13 \def Ì{\~A}                      
\catcode205=13 \def Í{\~O}                      
\catcode206=13 \def Î{\Theta}              
\catcode207=13 \def Ï{\theta}               
\catcode208=13 \def Ð{\relax\ifmmode\bar\else\expandafter\=\fi}
\catcode209=13 \def Ñ{\overline}             
\catcode210=13 \def Ò{\langle}               
\catcode211=13 \def Ó{\relax\ifmmode\{\else\ital\fi}      
\catcode212=13 \def Ô{\rangle}               
\catcode213=13 \def Õ{\}}                        
\catcode214=13 \def Ö{\sla}                      
\catcode215=13 \def ×{\relax\ifmmode\check\else\expandafter\v\fi}
\catcode216=13 \def Ø{\"y}                     
\catcode217=13 \def Ù{\"Y}  		    
\catcode218=13 \def Ú{\Leftarrow}       
\catcode219=13 \def Û{\Leftrightarrow}       
\catcode220=13 \def Ü{\relax\ifmmode\Rightarrow\else\sect\fi}
\catcode221=13 \def Ý{\sum}                  
\catcode222=13 \def Þ{\prod}                 
\catcode223=13 \def ß{\widehat}              
\catcode224=13 \def à{\pm}                     
\catcode225=13 \def á{\nabla}                
\catcode226=13 \def â{\quad}                 
\catcode227=13 \def ã{\in}               	
\catcode228=13 \def ä{\star}      	      
\catcode229=13 \def å{\sqrt}                   
\catcode230=13 \def æ{\^E}			
\catcode231=13 \def ç{\Upsilon}              
\catcode232=13 \def è{\"E}    	   	 
\catcode233=13 \def é{\`E}               	  
\catcode234=13 \def ê{\Sigma}                
\catcode235=13 \def ë{\Delta}                 
\catcode236=13 \def ì{\Phi}                     
\catcode237=13 \def í{\`I}        		   
\catcode238=13 \def î{\iota}        	     
\catcode239=13 \def ï{\Psi}                     
\catcode240=13 \def ð{\times}                  
\catcode241=13 \def ñ{\Lambda}             
\catcode242=13 \def ò{\cdots}                
\catcode243=13 \def ó{\^U}			
\catcode244=13 \def ô{\`U}    	              
\catcode245=13 \def õ{\bo}                       
\catcode246=13 \def ö{\relax\ifmmode\hat\else\expandafter\^\fi}
\catcode247=13 \def÷{\relax\ifmmode\tilde\else\expandafter\~\fi}
\catcode248=13 \def ø{\ll}                         
\catcode249=13 \def ù{\gg}                       
\catcode250=13 \def ú{\eta}                      
\catcode251=13 \def û{\kappa}                  
\catcode252=13 \def ü{\half}     		 
\catcode253=13 \def ý{\Gamma} 		
\catcode254=13 \def þ{\Xi}   			
\catcode255=13 \def ÿ{\relax\ifmmode{}^{\dagger}{}\else\dag\fi}


\def\ital#1Õ{{\it#1\/}}	     
\def\un#1{\relax\ifmmode\underline#1\else $\underline{\hbox{#1}}$
	\relax\fi}

\def\tdt#1{\on{\hbox{\bf .\kern-1pt.\kern-1pt.}}#1}   
\def\({\eqno(}

\def\refs{\sect{REFERENCES}\par\medskip \frenchspacing 
	\parskip=0pt \refrm \baselineskip=1.23em plus 1pt
	\def\ital##1Õ{{\refit##1\/}}}


\def\õ#1{
	\screwcount\num
	\num=1
	\screwdimen\downsy
	\downsy=-1.5ex
	\mkern-3.5mu
	õ
	\loop
	\ifnum\num<#1
	\llap{\raise\num\downsy\hbox{$õ$}}
	\advance\num by1
	\repeat}
\def\upõ#1#2{\screwcount\numup
	\numup=#1
	\advance\numup by-1
	\screwdimen\upsy
	\upsy=.75ex
	\mkern3.5mu
	\raise\numup\upsy\hbox{$#2$}}


\catcode`\|=\active \catcode`\<=\active \catcode`\>=\active 
\def|{\relax\ifmmode\delimiter"026A30C \else$\mathchar"026A$\fi}
\def<{\relax\ifmmode\mathchar"313C \else$\mathchar"313C$\fi}
\def>{\relax\ifmmode\mathchar"313E \else$\mathchar"313E$\fi}


\def\d{{\cal D}}
\def\n{{\cal N}}
\paper

ACTIONS FOR QCD-LIKE STRINGS

\author\WS\ITP

96-1

January 2, 1996	

We introduce a random lattice corresponding to ordinary Feynman
diagrams, with $1/p^2$ propagators instead of the Gaussians used in the
usual strings.  The continuum limit defines a new type of string action
with two worldsheet metrics, one Minkowskian and one Euclidean.  The
propagators correspond to curved lightlike paths with respect to the
Minkowskian worldsheet metric.  Spacetime dimensionality of four is
implied not only as the usual critical dimension of renormalizable quantum
field theory, but also from T-duality.

Ü1.  INTRODUCTION

Originally relativistic string theory was introduced to describe hadrons
directly.  With the advent of quantum chromodynamics the hadronic string
was identified as a bound state of gluons and quarks.  Unfortunately, the
available string models disagreed with several properties expected of QCD
strings:  (1) They required spacetime dimensions greater than four.  This
higher dimensionality is usually called the ``critical" dimension, and is
hidden by eliminating the extra dimensions through compactification or
related mechanisms.  However, there is no understanding as to why
exactly four dimensions remain uncompactified.  On the other hand,
relativistic quantum field theory of particles (QCD in particular) also has a
critical dimension:  The condition of perturbative renormalizability
requires such theories be defined in no more than four dimensions.  In this
sense particles, unlike strings, predict the correct dimensionality of
spacetime.  (An exception might be $Ä^3$ theory in six dimensions, but it
has an unbounded potential and lacks fermions.  A string analog to this
pathological theory is the bosonic string in 26 dimensions, which also has
a higher critical dimension than better behaved theories.)  (2) The
underlying partons (gluons and quarks) of hadrons are revealed in
processes with large transverse momenta (such as fixed-angle elastic
scattering), through the power-law behavior caused by the parton
propagators.  Unfortunately, the known string theories have Gaussian
dependence on these momenta in such limits [1].  (However, Green [2] has
found that power-law behavior can be obtained from the usual strings by
replacing Neumann boundary conditions with Dirchlet ones [3].)  This
behavior is closely related to the renormalizability of these string
theories in greater than four dimensions.  

These and other problems led to the reinterpretation of the known strings
as ``fundamental" strings, whose perturbative states were not to be
identified with hadrons but with gluons and quarks, as well as gravitons,
leptons, etc.  However, it would still be desirable for strings whose states
are identified with the fundamental particles to have critical dimension
four.  Thus, if it is possible to find a new kind of string theory that is
capable of describing hadrons, such a theory might also be preferable for
describing fundamental particles.  (Of course, there would still be at least
the distinction that some fundamental particles are massless, while all
hadrons are massive.)

One approach that helps to explain the differences between fundamental
strings and composite strings is to replace the worldsheet with a lattice
corresponding to a Feynman diagram of an underlying scalar field theory
[4].  By identifying the functional integration over worldsheet metrics in
the quadratic string action [5] with a sum over all lattices [6], and using
the large-N expansion  to define surfaces on these lattices [7], a
continuum limit can be defined for these lattice models that yields the
usual string (at least for the bosonic Liouville string in D$²$1, but formally
for all D) [8].  The key ingredient [4] is to note that the (worldsheet)
continuum limit of the product of Gaussian propagators in a Feynman
diagram is the exponential of the usual (mechanics) string action. 
Because Gaussian propagators ($e^{-p^2}$) are used, (1) all momentum
integrations are finite (removing any trace of the renormalizability
condition D=4), and (2) any parton behavior will give Gaussian behavior
rather than power-law.  (Usually it is argued that 
$e^{-p^2}®1/(1+p^2)$, but clearly this approximation cannot describe
massless particles such as gluons, and is also inaccurate for large
momenta.)

Our approach will be to modify the random lattice formulation of string
theory so that the underlying field theory has true $1/p^2$ propagators. 
In general, we then expect to be able to describe two kinds of strings:  (1)
In the case where the underlying theory is finite (such as $\n$=4
supersymmetric Yang-Mills theory), we expect a gravity string. 
Finiteness implies conformal invariance (in the absence of classical mass
terms), and even if it is spontaneously broken (generating mass) there is
at least a (massless) dilaton as its Goldstone boson.  If the conformal
graviton is among the string states, it can then eat the dilaton to become
the ordinary graviton [9]. (2)  In the case where the underlying field
theory is asymptotically free (such as QCD), we expect a hadronic string. 
As usual, the mass scale is generated by dimensional transmutation, and
all string states might be massive.  (On the other hand, the usual string
theories already have a mass scale classically, but also have massless
states.  However, in the random lattice approach the Liouville mode
apparently makes even these states massive.)  In this paper the theory
we consider explicitly is wrong-sign (massless) $Ä^4$ theory in four
dimensions, which is asymptotically free.  (The wrong sign for the
coupling, and thus asymptotic freedom, is a consequence of the random
lattice approach.) 

The outline of the paper is:  (a) In the next section we use Schwinger
parameters to exponentiate Feynman diagram integrands without
changing the propagators.  This produces a random lattice action whose
classical extrema for any particular Feynman diagram are its Landau
singularities.  Individual propagators now satisfy T-duality only in D=4
(and when massless).  (b) We then take the continuum limit for $Ä^4$
theory.  The action has the usual worldsheet metric, appearing in the
Virasoro conditions, but which is now Minkowskian, and a second metric
that is Euclidean and appears in the $x$ equations.  The four-point
vertices of the lattice have become the worldsheet light-cones with
respect to the Minkowskian metric.  (c) We discuss modifications of this
action that might be useful for more general theories.  (d) Finally, we
demonstrate that D=4 for the continuum string is implied by either
renormalizability (as for the lattice theory) or T-duality.

Ü2. FEYNMAN DIAGRAMS

The main reason for introducing Gaussian propagators was to allow
exponentiation of the integrand of the second-quantized Feynman
diagram, so that it could be identified with a (discretized) first-quantized
path integral.  The exponentiation of propagators is more commonly
performed through the use of Schwinger parameters,
$$ {1\over p^2} = Ç_0^¥ d ¼e^{- p^2} $$
 (For simplicity, we restrict ourselves to massless particles.) Even in this
discretized form, Feynman diagrams are very naturally associated with
first-quantized path integrals:  To make this analysis, we first write a
Feynman diagram for a scalar field theory with nonderivative
self-interactions in coordinate space as a Fourier transform:
$$ Ç dx'_i dp_{ij} d _{ij}¼e^{-Ý_{ÒijÔ}[ü _{ij}p_{ij}^2 +i(x_i-x_j)Ép_{ij}]} $$
 where $i,j$ label vertices (and external endpoints), $ÒijÔ$ labels links
(propagators), and $dx'_i$ integrations are over just vertices and not free
ends of external lines.  

Integration over these $x'$'s produces delta functions for momentum
conservation
$$ Ý_j p_{ij} = 0 $$
 for the propagator momenta $p_{ij}$.  The solution is to replace these
momenta with loop (and external) momenta:
$$ p_{ij} = p_{IJ} = k_I - k_J $$
 where $I,J$ label the loops; $p_{IJ}$ is now labeled by the two loops on
either side of the corresponding propagator, instead of the two vertices
at either end.  (This parametrization works for planar diagrams.  Our
arguments generalize straightforwardly for more complicated topologies,
but the explicit expressions are messier because source terms for external
lines modify the form of the momentum conservation law.)  The labelling
can be uniquely defined by the $1/N$ expansion:  Each loop is then defined
as a quark loop (including broken loops for external lines), associated with
the indices of the scalar field matrix (scalar = quark + antiquark); the loop
momentum
$k_I$ can then be thought of as the quark momentum.  Now external
momenta are also expressed as differences, so total momentum
conservation is automatic, and there is a momentum ``translation
invariance".  The Feynman integral becomes
$$ Ç dk'_I d _{IJ}¼e^{-Ý_{ÒIJÔ}ü _{IJ}(k_I -k_J)^2} $$
 This transformation from coordinate space to (loop-)momentum space is
the usual T-duality transformation [10]:  Starting from the ``first-order"
form with auxiliary variable $p_{ij}$, instead of integrating out $p$ we
integrated out $x$.  These are the same manipulations performed on the
usual string (as generalized from the continuum [11] to the random lattice
[12]), except there there are no $ $'s.  The variable dual to the coordinate
$x_i$ is the loop momentum $k_I$, so the duality transformation is a
Fourier transformation as well as a transformation to the dual lattice
(vertex $ª$ loop).  If we had included an external constant spacetime
metric, we would also have seen that, while $x_i-x_j$ is a contravariant
vector, $k_I-k_J$ is a covariant one (as usual for momenta).

Since we now interpret the exponent of the Feynman integrand as a
classical mechanics action, the natural thing to do is vary it to find the
classical equations of motion.  We first note that the $ $'s are constrained
to be positive:  We can either check the action at $ =0$ to see whether it
is an extremum, or make a change of variables such as $ =º^2$ or $e^º$
to an unconstrained variable.  The resulting equations are then (in
addition to the momentum conservation law, which we already solved)
$$  _{IJ}(k_I -k_J)^2 = 0,âÝ_J  _{IJ}(k_I -k_J) = 0 $$
 These are exactly the Landau equations for the singularities of
S-matrices (as derived from Feynman diagrams or from general properties
of S-matrix theory), which correspond to classical configurations of the
particles.  (The parameters $ $ in the Landau equations are often
identified with Feynman parameters rather than Schwinger parameters,
but our identification is more natural because (1) it is the Schwinger
parameters that correspond to the proper time, which is the classical
interpretation of the parameters in the Landau equations, (2) these
equations follow from the classical interpretation of our discretized
action, and (3) Feynman parameters also satisfy some irrelevant algebraic
conditions, while the Schwinger parameters are only constrained to be
positive, as usual for proper time.  The Feynman parameters can be
obtained as usual by scaling subsets of the Schwinger parameters and
integrating over the scaling parameters, leaving the Feynman parameters
as the remaining independent ones.)

Note that the Landau equations were obtained by variation of the
exponent for a single Feynman graph.  Since we are interpreting the
Feynman diagrams as the first-quantization of a string theory on a
random lattice, summation over all graphs replaces integration over the
worldsheet metric.  We have thus not performed variation with respect to
that metric, which is expected to show the stringy properties.  (In the
usual string theories, it gives the Virasoro conditions, which determine
the string spectrum; in our case the interpretation is not yet clear.)

The most important property of this duality transformation for the usual
strings is that the dual action is the same as the original action, only on
the dual lattice.  This is just the statement that the Gaussian propagator
takes the same form after Fourier transformation:
$$ e^{-k^2/2} ª e^{-x^2/2} $$
 This is not the case for conventional field theory propagators, even in the
massless case, ÓexceptÕ in four dimensions:
$$ {1\over k^2} ª ý(\f D2-1){1\over x^{D-2}} = {1\over x^2}
	\hbox{ in D=4} $$
 In terms of the discretized action, this is a consequence of spurious $ $
dependence of the measure from $p$ integration and from the required
transformation of $ $
$$   ª {1\over  } $$ 
 as easily seen from Fourier transformation of a single propagator written
in Schwinger parametrized form.  Since T-duality is a property of the
usual strings, this gives a strong hint that QCD-like strings prefer four
dimensions.

Ü3. CONTINUUM ACTIONS FOR $\hbox{\char'010}^{\displaystyle 4}$

We now look for continuum actions that will ÓexactlyÕ reproduce the
Feynman diagrams of ordinary field theory when the worldsheet is
replaced with a random lattice.  Such actions should represent the
continuum limit of conventional field theories.  The simplest way to derive
a continuum action from a random lattice action is to consider a regular
square (i.e., flat) lattice, where the continuum limit is easy to take, and
covariantize with respect to the worldsheet metric.  (This is the analog of
the procedure used to couple continuum theories to gravity.)  To this
continuum action we then add terms depending only on the worldsheet
metric, to give the relative weight of different graphs.  (Of course,
another continuum limit is to replace the propagators with
one-dimensional worldline actions and sew together the ends, but here
we look for two-dimensional continuous worldsheets.)  Since $x_i-x_j$
becomes the worldsheet vector $»x$, $p_{ij}$ becomes a vector (density),
so $ $ becomes a symmetric tensor (density).    For each vertex on a
regular square lattice there are two propagators, corresponding to the
two independent directions on the worldsheet.  This picture generalizes
straightforwardly to an arbitrary Feynman diagram in (wrong-sign) $Ä^4$
theory.  The continuum $ $ then has only two components at any point on
the worldsheet, so it must be a traceless symmetric tensor.  

If we impose the tracelessness condition through a Lagrange multiplier,
the continuum action on a curved worldsheet becomes
$$ L_1 = ü _{mn}(p^mÉp^n -Âg^{mn}) +ip^mÉ»_m x +L_g(g_{mn}) $$
 where
$$ L_g = å{-g}(µ -R¼ln¼N) $$
 contains the usual geometric factors:  The cosmological constant $µ$ on
the random lattice gives the coupling constant factor $e^{-µ}$ for each
vertex in the Feynman diagram, while the worldsheet curvature term
gives the string coupling $1/N$ of the topological expansion (at least for
its bare value).  For this form of the action all the functional integration
measures (except possibly that for $g_{mn}$) are the usual trivial ones;
integrating out $p^m$, for example, would introduce nontrivial $ $
dependence into the measure.  Note that $ $ acts as a metric independent
from $g$:  On the random lattice $g$ appears implicitly, defining the
geometry of the lattice, while $ $ appears in addition to that, and is
responsible for the $1/p^2$ propagators of the underlying field theory. 
(In [10,13] $ $ was identified with $g$ by approximating it with a
nonvanishing constant even at finite order in perturbation theory,
resulting in the usual string theory.)

We can also enforce the tracelessness condition through a gauge
invariance:  By solving the constraint on $ _{mn}$,
$$  _{mn} = ö _{mn} -üg_{mn}g^{pq}ö _{pq},â
	¶ö _{mn} = ½g_{mn}âÜ $$
$$ L_2 = üö _{mn}(p^mÉp^n -üg^{mn}g_{pq}p^pÉp^q) 
	+ip^mɻ_m x +L_g $$
 Furthermore, we can enforce positivity of $ _{mn}$ through the
redefinition
$$ ö _{mn} = º_m º_nâÜ $$
$$ L_3 = üº_m º_n (p^mÉp^n -üg^{mn}g_{pq}p^pÉp^q) 
	+ip^mɻ_m x +L_g $$
 (This change of variables also makes the functional integration measure
for $º$ nontrivial.)  

As usual, Weyl scale invariance is broken classically only by the
cosmological term, so the naive continuum limit is $µ£0$.  In the usual
bosonic string this becomes $µ£µ_c$ when quantum corrections are
taken into account, corresponding to a critical value for the $Ä^4$
coupling.  However, such conclusions might be modified here by
dimensional transmutation, since now the Feynman diagrams are not all
finite.  (On the other hand, in a model corresponding to the Feynman
diagrams of a finite supersymmetric Yang-Mills theory, $µ_c=0$ would
correspond to the value for S-self-duality.)  Also, if we work in the critical
dimension (four), we might expect that the string coupling $1/N$ is
unrenormalized.  (The continuum limit does not necessarily require $N£¥$
by definition, only the use of an expansion in $1/N$.)  

As for the usual Liouville approach to the random string, the classical
equations from varying $g_{mn}$ are not expected to make sense, but
need quantum corrections.  (Field redefinitions of $g_{mn}$ can cause
similar problems for the other fields.)  We might consider adding a term
$$ © R\f1õ R $$
 to $L_g$ to give $g$ nontrivial dynamics, for calculational purposes.  The
``bare" value of $©$ would then be taken to vanish, while a nonzero
coefficient could be generated quantum mechanically through a conformal
anomaly.

Since $ $ must be positive definite (as the Schwinger parameters were), it
is a Euclidean metric.  This implies that $g$ must be Minkowskian:  If we
perform a similarity transformation on $ $ to transform it into the
identity, then $g$ is traceless, and therefore Minkowskian.  (For the usual
random lattice and string, $g$ is Euclidean.)  The conformal gauge
corresponding to the random lattice is then the one where $g$ is
off-diagonal, so the constraint on $ $ forces it to be diagonal, with the
two diagonal components corresponding to the two lightlike directions
with respect to $g$.  (One could also choose a gauge where
$ _{mn}¾¶_{mn}$.  Since not both $g$ and $ $ can be fixed, there is always
a ``spin-2" field.)  Thus the edges of the $Ä^4$ lattice from which we
derived this continuum theory are lightlike paths on a Minkowskian
worldsheet; from any vertex Óthe four propagators form the light-coneÕ at
that point on the worldsheet.  (On the Feynman diagram lattice all
curvature is associated with the plaquettes/quark loops, relating to the
number of sides, while all vertices and edges/propagators are flat; on the
dual lattice, where all plaquettes are squares, all curvature is associated
with the vertices.)  This is a natural lattice for a Minkowskian metric;
e.g., in the case of a regular square lattice, the areas $ë§^+ë§^-$ of the
plaquettes are globally Lorentz invariant, being also the 2D metric.

It should also be possible to generalize this geometric picture to general
renormalizable (4D) theories:  Since all such couplings are four-point or
less, they can all be represented by a subset of the four lightlike
worldsheet directions leaving any vertex.  This only requires a mechanism
to suppress the propagators in some of these directions.

Finally, we can also replace $g$ with a zweibein and use it to flatten the
indices on $ $:  Since $g$ is Minkowskian, we write 
$g_{mn}=e_{(m}{}^+ e_{n)}{}^-$, and thus
$$ L_2' = üö _{àà}p_¦Ép_¦ +ip_àÉe_¦{}^m »_m x +L_g $$
 or
$$ L_3' = ü(º_à)^2 p_¦Ép_¦ +ip_àÉe_¦{}^m »_m x +L_g $$

Ü4. RELATED ACTIONS

The equations of motion from varying $ö _{mn}$ for $ö _{mn}±0$ are the
usual Virasoro conditions with respect to $g_{mn}$:
$$ p^mÉp^n -üg^{mn}g_{pq}p^pÉp^q = 0 $$
 This is the continuum limit of the Feynman diagram mass-shell condition
$p^2=0$.  If we restrict ourselves to $Ä^4$ theory, we thus have
$(e_m{}^à p^m)^2=0$ in the two worldsheet light-cone directions
$e_m{}^à$, and their opposites ($-e_m{}^à$), corresponding to the four
legs on the four-point vertex.  However, for higher-point vertices we
would have additional legs, and so require additional world-sheet
directions that are linear combinations of $e_m{}^+$ and $e_m{}^-$.  This
would imply the stronger constraint
$$ p^mÉp^n = 0 $$
 Unlike the Lagrangian with constrained $ _{mn}$, which allows only $Ä^4$
or lower-order interactions (since only four worldsheet directions are
lightlike), an unconstrained $ _{mn}$ would allow vertices with arbitrary
powers of $Ä$, as in the usual strings.  To guarantee renormalizability,
this would require that $Ä$ be dimensionless, so the interaction vertex
would require derivatives.  However, elimination of the $Âg^{mn} _{mn}$
term would (at least classically) completely decouple $g_{mn}$ from the
other variables; this might require recoupling through the additional
terms representing these vertex factors.  Some examples of such theories
are 2D nonlinear sigma models and 4D $\n$=1 super Yang-Mills theory in
superspace.  More generally, we have the dimensional analysis for a
kinetic term of the form (with $Ï$ counting appropriate to D=3,4,6,10)
$$ Çd^D xÊd^{2\n(D-2)}ϼüVõV $$
$$ Üâ0 = -D +\n(D-2) +0 +2 +0 = (\n-1)(D-2)âÜâ\hbox{$\n$=1 or D=2} $$

Another possible generalization would be to have $ _{mn}g^{mn}$ appear
differently from the other components, so it would not act as a Schwinger
parameter.  For example, if we add a term of the form
$$ ÂÐÏÏ $$
 to $L_1$, in terms of new anticommuting variables $Ï$ and $ÐÏ$, then in
the Feynman diagrams integrating over $Â,Ï,ÐÏ$ and $ _{mn}g^{mn}$
produces a momentum-dependent vertex, of the form $Ä^2(»Ä)^2$, since
$ _{mn}g^{mn}$ multiplies $p^m$'s in two different directions on the
worldsheet.  In the present case that would lower the critical dimension
to two.  However, in models with spin this might be a useful method to
introduce vertex factors.  Also, a term 
$$ e^{-µ}ÐÏÏ $$
 might be an alternative to the cosmological term for introducing the
$Ä^4$ coupling.

The Lagrangian $L_1$ is very similar to one introduced by Polyakov for the
usual bosonic string [14]:  In first-order form,
$$ L_P = ü _{mn}(p^mÉp^n +gg^{mn}) +ip^mÉ»_m x +L_g $$
 which differs from $L_1$ only by the substitution
$$ Â £ -g $$
 and dropping the positivity condition on $ _{mn}$.  Varying with respect
to the now-Lagrange-multiplier $ $ gives the Nambu-Goto action; on the
other hand, varying with respect to $g_{mn}$ gives
$$  _{mn} = µ(-g)^{-1/2}g_{mn} $$
 which results in the usual quadratic-in-$x$ (and $p$) string action (at
least classically) with the identification
$$ µ = Œ' $$
 In the usual random lattice interpretation, both $g$ and $ $ are
Euclidean in this case.

This suggests a perturbation expansion for $L_1$ based on introducing the
length scale through the vacuum value
$$ Ò(-g)^{-1}ÂÔ = Œ'^{-1} $$
 (Such a result might also be expected from one-loop calculations in the
usual perturbation expansion, in analogy to the CP(n) model.)  By
comparison with the Polyakov action, we see that perturbation about
$ÒÂÔ$ would then be perturbation about a conventional string theory, but
with the additional variable $Â$ (assuming $Â$ develops nontrivial
dynamics from loops, so it is no longer just a Lagrange multiplier).  This
would correspond to a different phase from $Ä^4$ theory, since in the
Polyakov action $g$ and $ $ have the same signature (which we would
now choose to be Minkowskian in line with our geometric picture for
renormalizable field theory).  Such an expansion might be suitable for
studying Regge behavior.

Ü5. D=4 FROM THE WORLDSHEET CONTINUUM

Eliminating $p^m$ from $L_1$ by functional integration yields
$$ L_x = ü( ^{-1})^{mn}(»_m x)É(»_n x) - _{mn}g^{mn} +L_g $$
 and introduces a factor of $ ^{-D/2}$ into the measure, where
$ ­det( _{mn})$.  On the other hand, eliminating $x$ (and $p^m$) by
solving 
$$ »_m p^m=0Üp^m=·^{mn}»_n k $$
 yields
$$ L_k = ü ( ^{-1})^{mn}(»_m k)É(»_n k) - _{mn}g^{mn} +L_g $$
 The duality transformation from $x£k$ is therefore an invariance of the
Lagrangian (classical) if we also transform
$$  _{mn} £  ^{-1} _{mn}â[=·_{mp}·_{nq}( ^{-1})^{pq}],â £  Â $$
 and an invariance of the functional measure (one-loop) if D=4: 
Compared to the measure factor $ ^{-D/2}$ for $L_x£ ^{+D/2}$ for $L_k$,
the $ _{mn}$ transformation on $\d  $ generates an additional factor
$ ^{-2}$ in the measure for $L_k$ from the two nonvanishing components
of $ _{mn}$ (or equivalently, $ ^{-3}$ from the change in the $ _{mn}$
measure together with $ ^{+1}$ from the change in the $Â$ measure). 
Thus, D=4 arises at one loop as the condition for cancellation of the
T-duality anomaly.  Note that T-duality is then an invariance of the
continuum theory; it was not an exact symmetry on the random lattice,
since duality replaced each lattice with its dual, which did not appear in
the original Feynman diagram summation (at least for $Ä^4$ theory). 
At least for the usual strings, T-duality is related to the standard
holomorphic properties, and therefore to conformal invariance.

While variation of these new actions with respect to $ _{mn}$ produces
the usual Virasoro conditions using the Minkowskian metric $g_{mn}$,
variation with respect to $x$ and $p$ results in equations for $x$ similar
to the usual except for the replacement of $(-g)^{-1/2}g_{mn}$, which has
determinant $-1$, with the Euclidean $ _{mn}$ (a density of the same
weight), which has as its determinant a nontrivial scalar.  Thus $ _{mn}$
acts as the metric for purposes of the $x$ equations of motion.  To
perform quantum calculations we need to give a vacuum expectation
value to $ _{mn}$ to define worldsheet propagators for $x$.  From 4D
dimensional analysis, we expect the expectation
$$ Ò _{mn}Ô = Œ' ¶_{mn} $$
 or covariantly
$$ Ò Ô = Œ'^2 $$
 Since we already have
$$ Òg_{mn}Ô = ú_{mn} ± ¶_{mn} $$
 this suggests a Higgs effect for 2D general coordinate invariance.  

To see the critical dimension following from renormalizability, we
perform the continuum version of the calculation used to find the
superficial divergence of Feynman diagrams.  Including source terms for
external momenta (or considering just planar graphs), the functional
integral for $L_k$ is of the form
$$ Ç\d^2  ¼\d^D k'¼e^{-k k} $$
 where all $§$-dependence and indices are implicit ($ $ no longer refers
to the determinant), $k$ includes the worldsheet variables $k'$ as well as
the external (source) momenta $q$, the Lagrange multiplier constraint has
been solved, and we have ignored dependence on and integration over
$g_{mn}$.  Performing the $k'$ integration leaves
$$ Ç\d^2  ¼ ^{-D/2}¼e^{-q q} $$
 where $ ^{-D/2}$ is the resulting functional determinant (nonlocal in
$§$).  We now consider integration over just the zero-mode of $ $: 
Writing
$$  _{mn} (§) = Œ' Œ_{mn} (§) $$
 where $Œ'$ is now a variable independent of the worldsheet coordinates
$§$, we have
$$ Ç\d^2  ¼ ^{-D/2}¼dŒ'¼(Œ')^{2-D/2}¼e^{-Œ'qŒq} $$
 where the $(Œ')^{2-D/2}$ factor in the measure is at every point on the
worldsheet (ignoring some finite power of $Œ'$ from the finite number of
external lines).  The $Œ'$ integration gives the usual $ý$-function factor
$$ ý[(2-\f D2)V] $$
 where $V$ is the infinite volume of the worldsheet, as measured by the
number of points on the worldsheet.  More precisely, the usual Feynman
diagram integration for $Ä^4$ theory gives
$$ ý[(2-\f D2)L -(2-\f E2)] $$
 where $L$ is the number of loops and $E$ is the number of external lines. 
For D>4 (nonrenormalizable) it gives $¥$ (when the argument is an
integer), for D<4 (superrenormalizable) the limit $L£¥$ also gives $¥$,
while for D=4 (renormalizable) the result is finite and nonvanishing for
$E>4$.  Like the duality argument, this suggests that the 4D theory is
better behaved than either D>4 or D<4.  In the continuum limit, where
$V$ is simply $¥$, the integration factor is divergent unless D=4 (and
sometimes even when D=4, depending on boundary contributions).  This
shows that the renormalizability condition D=4 for the critical
dimension survives the continuum limit.

ÜACKNOWLEDGMENTS

I thank Jan de Boer, Gordon Chalmers, Marty Halpern, Martin Ro×cek, Jack
Smith, and George Sterman for discussions. This work was supported in
part by the National Science Foundation Grant No.¼PHY 9309888.

\refs

£1 G. Veneziano, ÓNuo. Cim.Õ É57A (1968) 190;\\
	V. Alessandrini, D. Amati, and B. Morel, ÓNuo. Cim.Õ É7A (1971) 797;\\
	D.J. Gross and P.F. Mende, \PL 197B (1987) 129, \NP 303 (1988) 407;\\
	D.J. Gross and J.L. Ma÷nes, \NP 326 (1989) 73.
 
£2 M.B. Green, \PL 201B (1988) 42, É266B (1991) 325.

£3 W. Siegel, \NP 109 (1976) 244.

£4 H.B. Nielsen and P. Olesen, \PL 32B (1970) 203;\\
	D.B. Fairlie and H.B. Nielsen, \NP 20 (1970) 637;\\
	B. Sakita and M.A. Virasoro, \PR 24 (1970) 1146. 

£5 P.A. Collins and R.W. Tucker, \PL 64B (1976) 207;\\
	L. Brink, P. Di Vecchia, and P. Howe, \PL 65B (1976) 471;\\
	S. Deser and B. Zumino, \PL 65B (1976) 369.

£6 F. David, \NP 257 [FS14] (1985) 543;\\
	V.A. Kazakov, I.K. Kostov and A.A. Migdal, \PL 157B (1985) 295. 

£7 G. 't Hooft, \NP 72 (1974) 461. 

£8 M.R. Douglas and S.H. Shenker, \NP 335 (1990) 635;\\
	D.J. Gross and A.A. Migdal, \PR 64 (1990) 127;\\
	E. Br«ezin and V.A. Kazakov, \PL 236B (1990) 144. 

£9 F. Englert, C. Truffin, and R. Gastmans, \NP 117 (1976) 407.

£10 F. David and R. Hong Tuan, \PL 158B (1985) 435.

£11 W. Siegel, \PL 134 (1984) 318;\\
        T.H. Buscher, \PL 194B (1987) 59, É201B (1988) 466. 

£12 D.V. Boulatov, V.A. Kazakov, I.K. Kostov, and A.A. Migdal,
        \NP 275 [FS17] (1986) 641;\\
	W. Siegel, \PL 252B (1990) 558. 

£13 R. Hong Tuan, \PL 173B (1986) 279, É286 (1992) 315, 
	hep-th/9507106 (Orsay preprint).

£14 A.M. Polyakov, \NP 268 (1986) 406.
	
\bye